
\def\service{T}
\catcode`\@=11
\def\unredoffs{\voffset=11mm \hoffset=0.5mm}

%
\newbox\leftpage \newdimen\fullhsize \newdimen\hstitle \newdimen\hsbody
\newdimen\hdim
\tolerance=400\pretolerance=800
%
%
\newif\ifsmall \smallfalse
\newif\ifdraft \draftfalse
\newif\iffrench \frenchfalse
\newif\ifeqnumerosimple \eqnumerosimplefalse
\nopagenumbers
\headline={\ifnum\pageno=1\hfill\else\hfil{\headrm\folio}\hfil\fi}
\def\draftstart{
\magnification=1200 \unredoffs\hsize=130mm\vsize=190mm
\hsbody=\hsize \hstitle=\hsize 
\nolabels
\iffrench
\dicof
\else
\dicoa
\fi
}

\font\elevrm=cmr9

\newdimen\chapskip
\font\twbf=cmssbx10 scaled 1200
\font\ssbx=cmssbx10

\font\caprm=cmr9
\font\capit=cmti9
\font\capbf=cmbx9
\font\capsl=cmsl9
\font\capmi=cmmi9
\font\capex=cmex9
\font\capsy=cmsy9
\chapskip=17.5mm
\def\makeheadline{\vbox to 0pt{\vskip-22.5pt
\line{\vbox to8.5pt{}\the\headline}\vss}\nointerlineskip}
\font\tbfi=cmmib10
\font\tenbi=cmmib7
\font\fivebi=cmmib5
\textfont4=\tbfi
\scriptfont4=\tenbi
\scriptscriptfont4=\fivebi
\font\headrm=cmr10

\font\eightrm=cmr6
\font\sixrm=cmr5
\font\eightmi=cmmi6
\font\sixmi=cmmi5
\font\eightsy=cmsy6
\font\sixsy=cmsy5
\font\eightbf=cmbx6
\font\sixbf=cmbx5
\skewchar\capmi='177 \skewchar\eightmi='177 \skewchar\sixmi='177
\skewchar\capsy='60 \skewchar\eightsy='60 \skewchar\sixsy='60

\def\elevenpoint{
\textfont0=\caprm \scriptfont0=\eightrm \scriptscriptfont0=\sixrm
\def\rm{\fam0\caprm}
\textfont1=\capmi \scriptfont1=\eightmi \scriptscriptfont1=\sixmi
\textfont2=\capsy \scriptfont2=\eightsy \scriptscriptfont2=\sixsy
\textfont3=\capex \scriptfont3=\capex \scriptscriptfont3=\capex
\textfont\itfam=\capit \def\it{\fam\itfam\capit} 
\textfont\slfam=\capsl  \def\sl{\fam\slfam\capsl} 
\textfont\bffam=\capbf \scriptfont\bffam=\eightbf
\scriptscriptfont\bffam=\sixbf
\def\bf{\fam\bffam\capbf} 
\textfont4=\tbfi \scriptfont4=\tenbi \scriptscriptfont4=\tenbi
\normalbaselineskip=13pt
\setbox\strutbox=\hbox{\vrule height9.5pt depth3.9pt width0pt}
\let\big=\elevenbig \normalbaselines \rm}

\catcode`\@=11

\font\tenmsa=msam10
\font\sevenmsa=msam7
\font\fivemsa=msam5
\font\tenmsb=msbm10
\font\sevenmsb=msbm7
\font\fivemsb=msbm5
\newfam\msafam
\newfam\msbfam
\textfont\msafam=\tenmsa  \scriptfont\msafam=\sevenmsa
  \scriptscriptfont\msafam=\fivemsa
\textfont\msbfam=\tenmsb  \scriptfont\msbfam=\sevenmsb
  \scriptscriptfont\msbfam=\fivemsb

\def\hexnumber@#1{\ifcase#1 0\or1\or2\or3\or4\or5\or6\or7\or8\or9\or
	A\or B\or C\or D\or E\or F\fi }

\font\teneuf=eufm10
\font\seveneuf=eufm7
\font\fiveeuf=eufm5
\newfam\euffam
\textfont\euffam=\teneuf
\scriptfont\euffam=\seveneuf
\scriptscriptfont\euffam=\fiveeuf
\def\frak{\ifmmode\let\next\frak@\else
 \def\next{\Err@{Use \string\frak\space only in math mode}}\fi\next}
\def\goth{\ifmmode\let\next\frak@\else
 \def\next{\Err@{Use \string\goth\space only in math mode}}\fi\next}
\def\frak@#1{{\frak@@{#1}}}
\def\frak@@#1{\fam\euffam#1}

\edef\msa@{\hexnumber@\msafam}
\edef\msb@{\hexnumber@\msbfam}

\def\Bbb{\ifmmode\let\next\Bbb@\else
 \def\next{\errmessage{Use \string\Bbb\space only in math mode}}\fi\next}
\def\Bbb@#1{{\Bbb@@{#1}}}
\def\Bbb@@#1{\fam\msbfam#1}

\catcode`\@=12
\def\sla#1{\mkern-1.5mu\raise0.4pt\hbox{$\not$}\mkern1.2mu #1\mkern 0.7mu}
\def\Dbar{\mkern-1.5mu\raise0.4pt\hbox{$\not$}\mkern-.1mu {\rm D}\mkern.1mu}
\def\Abar{\mkern1.mu\raise0.4pt\hbox{$\not$}\mkern-1.3mu A\mkern.1mu}
\def\dicof{
\gdef\Resume{RESUME}
\gdef\Toc{Table des mati\`eres}
\gdef\soumisa{Soumis \`a:}
}
\def\dicoa{
\gdef\Resume{ABSTRACT}
\gdef\Toc{Table of Contents}
\gdef\soumisa{Submitted to}
}

\def\uniset{\rlap{\elevrm 1}\kern.15em 1}
\def\bkR{{\rm I\kern-.17em R}}
\def\bkC{{\rm \kern.24em
            \vrule width.05em height1.4ex depth-.05ex
            \kern-.26em C}}

\def\frac#1#2{{\textstyle{#1\over#2}}}

\def\leaderfill{\leaders\hbox to 1em{\hss.\hss}\hfill}
\def\saclay{\if S\service \spec \else \spht \fi}
\def\spht{
\centerline{Service de Physique Th\'eorique, CEA-Saclay}
\centerline{F-91191 Gif-sur-Yvette Cedex, FRANCE}}
\def\spec{
\centerline{Service de Physique de l'Etat Condens\'e, CEA-Saclay}
\centerline{F-91191 Gif-sur-Yvette Cedex, FRANCE}}
\def\logo{
\if S\service 
\font\sstw=cmss10 scaled 1200
\font\ssx=cmss8
\vtop{\hsize 9cm
{\sstw {\twbf P}hysique de l'{\twbf E}tat {\twbf C}ondens\'e \par}
\ssx SPEC -- DRECAM -- DSM\par
\vskip 0.5mm
\sstw CEA -- Saclay \par
}
\else 
\vtop{\hsize 9cm
}
\fi }
\catcode`\@=11
\def\deqalignno#1{\displ@y\tabskip\centering \halign to
\displaywidth{\hfil$\displaystyle{##}$\tabskip0pt&$\displaystyle{{}##}$
\hfil\tabskip0pt &\quad
\hfil$\displaystyle{##}$\tabskip0pt&$\displaystyle{{}##}$
\hfil\tabskip\centering& \llap{$##$}\tabskip0pt \crcr #1 \crcr}}
\def\deqalign#1{\null\,\vcenter{\openup\jot\m@th\ialign{
\strut\hfil$\displaystyle{##}$&$\displaystyle{{}##}$\hfil
&&\quad\strut\hfil$\displaystyle{##}$&$\displaystyle{{}##}$
\hfil\crcr#1\crcr}}\,}
\openin 1=\jobname.sym
\ifeof 1\closein1\message{<< (\jobname.sym DOES NOT EXIST) >>}\else%
\input\jobname.sym\closein 1\fi
\newcount\nosection
\newcount\nosubsection
\newcount\neqno
\newcount\notenumber
\newcount\figno
\newcount\tabno
\def\content{\jobname.toc}
\def\symbols{\jobname.sym}
\newwrite\toc
\newwrite\sym
\def\authorname#1{\centerline{\bf #1}\smallskip}
\def\address#1{ #1\medskip}
\newdimen\hulp
\def\maketitle#1{
\edef\oneliner##1{\centerline{##1}}
\edef\twoliner##1{\vbox{\parindent=0pt\leftskip=0pt plus 1fill\rightskip=0pt
plus 1fill
                     \parfillskip=0pt\relax##1}}
\setbox0=\vbox{#1}\hulp=0.5\hsize
                 \ifdim\wd0<\hulp\oneliner{#1}\else
                 \twoliner{#1}\fi}
\def\pacs#1{{\bf PACS numbers:} #1\par}
\def\submitted#1{{\it {\soumisa} #1}\par}
\def\title#1{\gdef\titlename{#1}
\maketitle{
\twbf
{\titlename}}
\vskip3truemm\vfill
\nosection=0
\neqno=0
\notenumber=0
\figno=1
\tabno=1
\def\prefix{}
\def\eqprefix{}
\mark{\the\nosection}
\message{#1}
\immediate\openout\sym=\symbols
}
\def\preprint#1{\vglue-10mm
\line{ \logo \hfill {#1} }\vglue 20mm\vfill}
\def\abstract{\vfill\centerline{\Resume} \smallskip \begingroup\narrower
\elevenpoint\baselineskip10pt}
\def\endabstract{\par\endgroup \bigskip}
\def\mktoc{\centerline{\bf \Toc} \medskip\caprm
\parindent=2em
\openin 1=\jobname.toc
\ifeof 1\closein1\message{<< (\jobname.toc DOES NOT EXIST. TeX again)>>}%
\else\input\jobname.toc\closein 1\fi
 \bigskip}
\def\section#1\par{\vskip0pt plus.1\vsize\penalty-100\vskip0pt plus-.1
\vsize\bigskip\vskip\parskip
\message{ #1}
\ifnum\nosection=0\immediate\openout\toc=\content%
\edef\ecrire{\write\toc{\par\noindent{\ssbx\ \titlename}
\string\leaderfill{\noexpand\number\pageno}}}\ecrire\fi
\advance\nosection by 1\nosubsection=0
\ifeqnumerosimple
\else \xdef\eqprefix{\prefix\the\nosection.}\neqno=0\fi
\vbox{\noindent\bf\prefix\the\nosection\ #1}
\mark{\the\nosection}\bigskip\noindent
\xdef\ecrire{\write\toc{\string\par\string\item{\prefix\the\nosection}
#1
\string\leaderfill {\noexpand\number\pageno}}}\ecrire}

\def\appendix#1#2\par{\bigbreak\nosection=0
\notenumber=0
\neqno=0
\def\prefix{A}
\mark{\the\nosection}
\message{\appendixname}
\leftline{\ssbx APPENDIX}
\leftline{\ssbx\uppercase\expandafter{#1}}
\leftline{\ssbx\uppercase\expandafter{#2}}
\bigskip\noindent\nonfrenchspacing
\edef\ecrire{\write\toc{\par\noindent{{\ssbx A}\
{\ssbx#1\ #2}}\string\leaderfill{\noexpand\number\pageno}}}\ecrire}%

\def\subsection#1\par {\vskip0pt plus.05\vsize\penalty-100\vskip0pt
plus-.05\vsize\bigskip\vskip\parskip\advance\nosubsection by 1
\vbox{\noindent\it\prefix\the\nosection.\the\nosubsection\
\it #1}\smallskip\noindent
\edef\ecrire{\write\toc{\string\par\string\itemitem
{\prefix\the\nosection.\the\nosubsection} {#1}
\string\leaderfill{\noexpand\number\pageno}}}\ecrire
}
\def\note #1{\advance\notenumber by 1
\footnote{$^{\the\notenumber}$}{\sevenrm #1}}

\def\nolabels{\def\wrlabel##1{}\def\eqlabel##1{}\def\reflabel##1{}}
\def\writelabels{\def\wrlabel##1{\leavevmode\vadjust{\rlap{\smash%
{\line{{\escapechar=` \hfill\rlap{\sevenrm\hskip.03in\string##1}}}}}}}%
\def\eqlabel##1{{\escapechar-1\rlap{\sevenrm\hskip.05in\string##1}}}%
\def\reflabel##1{\noexpand\llap{\noexpand\sevenrm\string\string\string##1}}}
\global\newcount\refno \global\refno=1
\newwrite\rfile
\def\ref{[\the\refno]\nref}
\def\nref#1{\xdef#1{[\the\refno]}\writedef{#1\leftbracket#1}%
\ifnum\refno=1\immediate\openout\rfile=\jobname.ref\fi
\global\advance\refno by1\chardef\wfile=\rfile\immediate
\write\rfile{\noexpand\item{#1\ }\reflabel{#1\hskip.31in}\pctsign}\findarg}
\def\findarg#1#{\begingroup\obeylines\newlinechar=`\^^M\pass@rg}
{\obeylines\gdef\pass@rg#1{\writ@line\relax #1^^M\hbox{}^^M}%
\gdef\writ@line#1^^M{\expandafter\toks0\expandafter{\striprel@x #1}%
\edef\next{\the\toks0}\ifx\next\em@rk\let\next=\endgroup\else\ifx\next\empty%
\else\immediate\write\wfile{\the\toks0}\fi\let\next=\writ@line\fi\next\relax}}
\def\striprel@x#1{}
\def\em@rk{\hbox{}}

\def\addref#1{\immediate\write\rfile{\noexpand\item{}#1}} 
\def\listrefs{
\ifnum\refno=1 \else
\immediate\closeout\rfile\writestoppt\baselineskip=14pt%
\vskip0pt plus.1\vsize\penalty-100\vskip0pt plus-.1
\vsize\bigskip\vskip\parskip\centerline{{\bf References}}\bigskip%
{\frenchspacing%
\parindent=20pt\escapechar=` \input \jobname.ref\vfill\eject}%
\nonfrenchspacing
\fi}
\def\startrefs#1{\immediate\openout\rfile=\jobname.ref\refno=#1}
\def\xref{\expandafter\xr@f}\def\xr@f[#1]{#1}
\def\refs#1{[\r@fs #1{\hbox{}}]}
\def\r@fs#1{\ifx\und@fined#1\message{reflabel \string#1 is undefined.}%
\xdef#1{(?.?)}\fi \edef\next{#1}\ifx\next\em@rk\def\next{}%
\else\ifx\next#1\xref#1\else#1\fi\let\next=\r@fs\fi\next}
%
\newwrite\lfile
{\escapechar-1\xdef\pctsign{\string\%}\xdef\leftbracket{\string\{}
\xdef\rightbracket{\string\}}}

\def\writestop{\def\writestoppt{\immediate\write\lfile{\string\pageno%
\the\pageno\string\startrefs\leftbracket\the\refno\rightbracket%
\string\def\string\secsym\leftbracket\secsym\rightbracket%
\string\secno\the\secno\string\meqno\the\meqno}\immediate\closeout\lfile}}
\def\writestoppt{}\def\writedef#1{}
\def\eqnn{\global\advance\neqno by 1 \ifinner\relax\else%
\eqno\fi(\eqprefix\the\neqno)}
%
\def\eqnd#1{\global\advance\neqno by 1 \ifinner\relax\else%
\eqno\fi(\eqprefix\the\neqno)\eqlabel#1
{\xdef#1{($\eqprefix\the\neqno$)}}
\edef\ewrite{\write\sym{\string\def\string#1{($\eqprefix%
\the\neqno$)}}%
}\ewrite%
}
%
\def\eqna#1{\wrlabel#1\global\advance\neqno by1
{\xdef #1##1{\hbox{$(\eqprefix\the\neqno##1)$}}}
\edef\ewrite{\write\sym{\string\def\string#1{($\eqprefix%
\the\neqno$)}}%
}\ewrite%
}
\def\em@rk{\hbox{}}
\def\xeqn{\expandafter\xe@n}\def\xe@n(#1){#1}
\def\xeqna#1{\expandafter\xe@na#1}\def\xe@na\hbox#1{\xe@nap #1}
\def\xe@nap$(#1)${\hbox{$#1$}}
\def\eqns#1{(\e@ns #1{\hbox{}})}
\def\e@ns#1{\ifx\und@fined#1\message{eqnlabel \string#1 is undefined.}%
\xdef#1{(?.?)}\fi \edef\next{#1}\ifx\next\em@rk\def\next{}%
\else\ifx\next#1\xeqn#1\else\def\n@xt{#1}\ifx\n@xt\next#1\else\xeqna#1\fi
\fi\let\next=\e@ns\fi\next}
\def\fig{fig.~\the\figno\nfig}
\def\nfig#1{\xdef#1{\the\figno}%
\immediate\write\sym{\string\def\string#1{\the\figno}}%
\global\advance\figno by1}%
\def\xfig{\expandafter\xf@g}\def\xf@g fig.\penalty\@M\ {}%
\def\figs#1{figs.~\f@gs #1{\hbox{}}}%
\def\f@gs#1{\edef\next{#1}\ifx\next\em@rk\def\next{}\else%
\ifx\next#1\xfig #1\else#1\fi\let\next=\f@gs\fi\next}%
\long\def\figure#1#2#3{\midinsert
#2\par
{\elevenpoint
\setbox1=\hbox{#3}
\ifdim\wd1=0pt\centerline{{\bf Figure\ #1}\hskip7.5mm}%
\else\setbox0=\hbox{{\bf Figure #1}\quad#3\hskip7mm}
\ifdim\wd0>\hsize{\narrower\noindent\unhbox0\par}\else\centerline{\box0}\fi
\fi}
\wrlabel#1\par
\endinsert}
\def\tab{table~\uppercase\expandafter{\romannumeral\the\tabno}\ntab}
\def\ntab#1{\xdef#1{\the\tabno}
\immediate\write\sym{\string\def\string#1{\the\tabno}}
\global\advance\tabno by1}
\long\def\table#1#2#3{\topinsert
#2\par
{\elevenpoint
\setbox1=\hbox{#3}
\ifdim\wd1=0pt\centerline{{\bf Table
\uppercase\expandafter{\romannumeral#1}}\hskip7.5mm}%
\else\setbox0=\hbox{{\bf Table
\uppercase\expandafter{\romannumeral#1}}\quad#3\hskip7mm}
\ifdim\wd0>\hsize{\narrower\noindent\unhbox0\par}\else\centerline{\box0}\fi
\fi}
\wrlabel#1\par
\endinsert}
\catcode`@=12
\def\draftend{\immediate\closeout\sym\immediate\closeout\toc
}
\draftstart
\preprint{T93/083}
\title{Exact scaling form for the collapsed 2D polymer phase}
\authorname{Bertrand Duplantier}
\address{\saclay}
\abstract
It has been recently argued that interacting self-avoiding walks
(ISAW)
of length $ \ell , $ in their low temperature phase (i.e. below the $ \Theta
$-point) should
have a partition function of the form:
$$ Q_{\ell}  \sim \mu^{ \ell}_ 0\mu^{ \ell^ \sigma}_ 1\ell^{ \gamma -1}\ ,
\eqno  $$
where $ \mu_ 0(T) $ and $ \mu_ 1(T) $ are respectively bulk and
perimeter monomer fugacities, both depending on the temperature $ T. $ In $ d
$
dimensions the
exponent $ \sigma $ could be close to $ (d-1)/d, $ corresponding to a $ (d-1)
$-dimensional
interface, while the configuration exponent $ \gamma $ should be universal in
the
whole collapsed phase. This was supported by a numerical study of 2D partially
{\sl directed\/} SAWs for which $ \sigma \simeq 1/2 $ was found.
I point out here that formula (1) already appeared at several places
in the two-dimensional case for which $ \sigma =1/2, $ and for which one can
even
conjecture the exact value of $ \gamma . $
\endabstract
\vfill
\submitted{Physical Review Letters (Comment)}
\eject
\eject
\magnification=1200
\baselineskip=18pt
It has been recently argued $\lbrack$1$\rbrack$ that interacting self-avoiding
walks (ISAW)
of length $ \ell , $ in their low temperature phase (i.e. below the $ \Theta
$-point) should
have a partition function of the form:
$$ Q_{\ell}  \sim \mu^{ \ell}_ 0\mu^{ \ell^ \sigma}_ 1\ell^{ \gamma -1}\ ,
\eqno (1) $$
where $ \mu_ 0(T) $ and $ \mu_ 1(T) $ are respectively bulk and
perimeter monomer fugacities, both depending on the temperature $ T. $ In $ d
$
dimensions the
exponent $ \sigma $ could be close to $ (d-1)/d, $ corresponding to a $ (d-1)
$-dimensional
interface, while the configuration exponent $ \gamma $ should be universal in
the
whole collapsed phase. This was supported by a numerical study of 2D partially
{\sl directed\/} SAWs for which $ \sigma \simeq 1/2 $ was found
$\lbrack$1$\rbrack$.
I point out here that formula (1) already appeared at several places
$\lbrack$2,3$\rbrack$
in the two-dimensional case for which $ \sigma =1/2, $ and for which one can
even
conjecture the exact value of $ \gamma . $

Dense 2D polymers filling a finite fraction $ f>0 $ of a lattice have indeed
been
studied in detail in $\lbrack$2$\rbrack$. The following formulae were
conjectured for the
numbers of configurations $ w^D_{\ell} $ of a dense {\sl linear\/} chain, and
$ w^D_{0,\ell} $ of a dense {\sl loop\/}
$$ \eqalignno{ w^D_{\ell} &  \sim  \left[\mu^ D(f) \right]^{\ell} {\rm
e}^{-B(f) \sqrt{ \ell}}  \ell^{\bar \gamma -1}\ , & (2) \cr w^D_{0,\ell} &
\sim  \left[\mu^ D(f) \right]^{\ell} {\rm e}^{-B(f) \sqrt{ \ell}}  \ell^{\bar
\gamma_ 0-1}\ , & (3) \cr} $$
where the bulk fugacity $ \mu^ D(f) $ and the perimeter free energy $ B(f) $
$\lbrack$4$\rbrack$ depend
on the filling fraction $ f. $ This is just (1) where $ \sigma =1/2 $ and $
\gamma =\bar \gamma , $ and where the
temperature $ T $ simply drives an effective value of $ f $ in the ISAW model.
Moreover, the {\sl ratio\/} $ w^D_{\ell} /w^D_{0,\ell} \sim \ell^{ \gamma^ D}
$ has been argued to be {\sl universal\/} and governed by
the exponent $ \gamma^ D \equiv  \bar \gamma  - \bar \gamma_ 0 = 19/16 $
given by a conformal field theory with a central charge $ c=-2 $
$\lbrack$2$\rbrack$. Formula
(3) was explicitly calculated for Hamiltonian walks on the Manhattan
lattice $\lbrack$3$\rbrack$ (i.e. for $ f=1), $ as well as the analogue of (2)
for
corner-to-corner walks, yielding the surface exponent $ \bar \gamma_{ 11} $
and its $ c=-2 $
universal part $ \gamma^ D_{11}\equiv\bar \gamma_{ 11}-\bar \gamma_ 0=5/8 $
$\lbrack$3$\rbrack$.

The value of the loop exponent $ \bar \gamma_ 0 $ itself (hence $ \bar \gamma
) $ does depend on the
{\sl boundary\/} conditions (periodic or free) imposed to the dense walk, and
in the
latter case also of the
{\sl shape\/} of the boundary domain $\lbrack$3$\rbrack$.
The case of Ref.$\lbrack$1$\rbrack$ clearly corresponds to {\sl free boundary
conditions\/}. In
Ref.$\lbrack$3$\rbrack$, the value of $ \bar \gamma_ 0 $ was given for a
boundary made of $ N $
wedges $ i $ of angles $ \alpha_ i, $ and separated by $ N $ {\sl smooth\/}
arcs $ \Gamma_ j $ of local curvature
radius $ \rho , $
$$ \eqalignno{\bar \gamma_ 0-1 & = -\zeta( 0) & (4) \cr \zeta( 0) & = \sum^
N_{i=1}{1 \over 24} \left({\pi \over \alpha_ i} - {\alpha_ i \over \pi}
\right) + \sum^ N_{j=1}{1 \over 12\pi}  \int^{ }_{ \Gamma_ j}{ {\rm d} s \over
\rho} \ , & (5) \cr} $$
a result taken from the spectral theory of the Dirichlet Laplacian
$\lbrack$5$\rbrack$.

When one lets the walk collapse by lowering $ T, $ as in $\lbrack$1$\rbrack$,
no perimeter shape
is imposed, but admitting in 2D $ \sigma =1/2 $ actually amounts to a
perimeter with
Hausdorff dimension 1. A
number $ N \longrightarrow \infty $ of smooth wedges in the walks boundary,
i.e.
a smooth perimeter, then seems the most likely.
In Eq.(5), the smooth arc term is the {\sl continuous\/} limit of the discrete
wedge
term. Thus, when $ N \longrightarrow \infty , $ $ \zeta( 0) $ reaches a
universal
value $ {1 \over 12\pi} \times 2\pi  = 1/6, $ valid for {\sl any\/} simple
smooth perimeter curve enclosing the walk.
It can be shown that this value also bounds (5) from below,
yielding the {\sl maximal\/} value $ \bar \gamma_ 0 = 5/6. $ This
substantiates our conjecture that
collapsed ISAW would have predominantly smooth perimeters, with loop and open
chain exponents
$$ \bar \gamma_ 0=5/6\ ,\ \ \ \gamma =\bar \gamma_ 0+\gamma^ D=97/48, $$
for which a numerical check would be welcome.

Notice that at $ T=0 $ exactly, the polymer phase should be Hamiltonian and
might
depend crucially on the minimal polygonal energy adapted to the lattice, and
the discrete version of (5) could again apply.

Finally, the value $ \gamma_{{\cal G}} $ for any collapsed network $ {\cal G}
$ is then obtained from $ \bar \gamma_ 0=5/6 $
and the theory developed in $\lbrack$3,2$\rbrack$.
\vglue 2truecm

\noindent {\bf References}

\item{$\lbrack$1$\rbrack$}A.L. Owczarek, T. Prellberg, R. Brak, {\sl Phys.
Rev. Lett.\/} {\bf 70}, 951
(1993).

\item{$\lbrack$2$\rbrack$}B. Duplantier, H. Saleur, {\sl Nucl. Phys. \/}{\bf
B290} $\lbrack$FS20$\rbrack$ 291 (1987).

\item{$\lbrack$3$\rbrack$}B. Duplantier, F. David, {\sl J. Stat. Phys.\/}
{\bf 51}, 327 (1988).

\item{$\lbrack$4$\rbrack$}A. Malakis, {\sl Physica \/}{\bf 84A}, 256 (1976).

\item{$\lbrack$5$\rbrack$}H.P. McKean, I.M. Singer, {\sl J. Diff. Geom.\/}
{\bf 1}, 43 (1967).

\listrefs
\draftend
\end